\documentclass[aps,prl,twocolumn,showpacs]{revtex4}

\usepackage{graphicx}
\usepackage{dcolumn}
\usepackage{amsmath}
\usepackage{wasysym}
\begin{document}

\title{Metal-insulator quantum critical point beneath the high $T_{\rm c}$ superconducting dome}

\author{Suchitra~E.~Sebastian$^1$}
\email{suchitra@phy.cam.ac.uk}
\author{N.~Harrison$^2$}
\email{nharrison@lanl.gov}
\author{M.~M.~Altarawneh$^2$}
\author{C.~H.~Mielke$^2$}
\author{Ruixing Liang$^{3,4}$}
\author{D.~A.~Bonn$^{3,4}$}
\author{W.~N.~Hardy$^{3,4}$}
\author{G.~G.~Lonzarich$^1$}

\affiliation{
$^1$Cavendish Laboratory, Cambridge University, JJ Thomson Avenue, Cambridge CB3~OHE, U.K\\
$^2$National High Magnetic Field Laboratory, LANL, Los Alamos, NM 87545\\
$^3$Department of Physics and Astronomy, University of British Columbia, Vancouver V6T 1Z4, Canada\\
$^4$Canadian Institute for Advanced Research, Toronto M5G 1Z8, Canada
}
\date{\today}

\begin{abstract}

An enduring question in correlated systems concerns whether superconductivity is favoured at a quantum critical point (QCP) characterised by a divergent quasiparticle effective mass. Despite such a scenario being widely postulated in high $T_{\rm c}$ cuprates and invoked to explain non-Fermi liquid transport signatures, experimental evidence is lacking for a critical divergence under the superconducting dome. We use ultra-strong magnetic fields to measure quantum oscillations in underdoped YBa$_2$Cu$_3$O$_{6+\rm x}$, revealing a dramatic doping-dependent upturn in quasiparticle effective mass at a critical metal-insulator transition beneath the superconducting dome. Given the location of this QCP under a plateau in $T_{\rm c}$ in addition to a postulated QCP at optimal doping, we discuss the intriguing possibility of two intersecting superconducting subdomes, each centred at a critical Fermi surface instability.

\end{abstract}
\maketitle

A continuous zero temperature instability between different groundstates$-$ termed as a quantum critical point$-$ is characterised by a divergence in a relevant susceptibility~\cite{millis1}. In strongly correlated systems~\cite{gegenwart1}, the influence of criticality on the entire body of itinerant electrons results in a global divergence of the effective mass$-$ which is recognized as the key defining experimental signature of quantum criticality~\cite{gegenwart1,coleman1}. The growth of electronic correlations on the zero temperature approach to the critical instability can be experimentally accessed by the tuning of parameters such as pressure and doping. Quantum oscillation measurements are ideally suited to investigate the effects of such tuning due to the direct access they provide to the effective mass of the elementary fermionic excitations that can be traced across the Quantum Critical Point (QCP)~\cite{shishido1}. Such a direct probe is crucial in superconducting materials, where bulk thermodynamic signatures of quantum critical behaviour of the normal quasiparticles~\cite{monthoux1,gegenwart1,millis1} are difficult to access due to the overlying superconducting dome.

While the emergence of high $T_{\rm c}$ superconductivity in the cuprate family is inextricably linked to the parent Mott insulating compound, remarkably little is known about the physics of the metal-insulator crossover~\cite{imada1} and its relation to electronic correlations. By using quantum oscillation measurements in strong magnetic fields to access normal state quasiparticles in underdoped YBa$_2$Cu$_3$O$_{6+x}$, we uncover a striking doping-dependent upturn in the effective mass at the location of the metal-insulator crossover~\cite{sun1,sonier1,taillefer1,li1,ando1}. Our findings provide bulk thermodynamic evidence for a metal-insulator quantum critical point (QCP) in high $T_{\rm c}$ cuprates~\cite{broun1,tallon1,chakravarty1,kivelson1,varma1,zaanen1,sachdev1}, without requiring extrapolation below the superconducting dome. The effective mass divergence unaccompanied by a change in Fermi surface area away from half-filling signals a novel many-body mechanism~\cite{brinkman1} that drives insulating behaviour in underdoped cuprates.

\begin{figure}
\centering 
\includegraphics*[width=.35\textwidth]{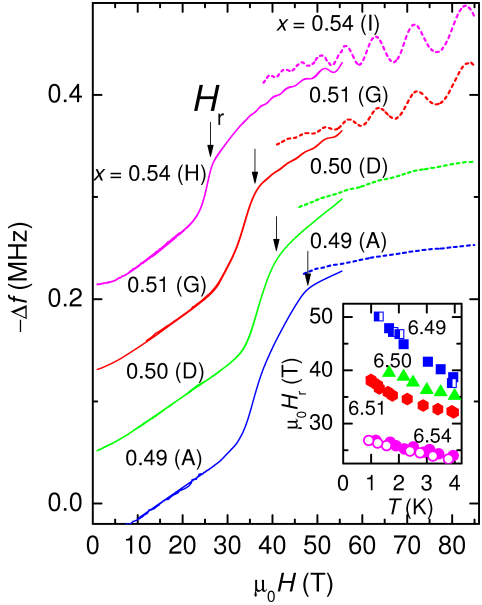}
\caption{An example of the measured magnetic field-dependent resonance frequency change $\Delta f$ (see text) for one sample of each composition at ~ 1.5 $-$ 1.6~K measured in both 55.5~T (shown by solid lines) and 85~T (shown by dotted lines) magnets~(see Appendix). The corresponding sample (i.e., A, D, G, H, I) is indicated in each case. Single crystalline platelets of average dimensions $\sim$~800~$\times$~500~$\times$~50~$\mu$m are coupled inductively to the face of a coil of $\sim$~5 turns (with its axis parallel to $H$) that forms part of the contactless conductivity circuit~(see Appendix). $H$ causes a crossover into a high magnetic field resistive state whose in-plane skin depth ($\sim$~100~$\mu$m at 46 MHz) increases the coil inductance causing $f$ to drop. The resistive crossover field~$H_{\rm r}$ (also plotted versus $T$ using solid symbols in the inset) is determined from the maximum in the derivative $\partial^2f/\partial B^2$. Hollow circles show static field measurements of $H_{\rm r}$ measured on a sample of one of the same compositions ($x=$~0.54) in a 45~T static field magnet. Given the thermal energy generated by flux flow motion in Type II superconductors, we ensure sample immersion in the liquid cryogen to ensure heat dissipation and well controlled temperatures~(see Appendix).}
\label{oscillations1}
\end{figure}

We trace the doping dependence of quantum oscillations with increased underdoping of YBa$_2$Cu$_3$O$_{6+x}$ ($x$=0.54, 0.51, 0.50, 0.49). Of the multiple Fermi surface orbits detected in a subset of samples of $x=$0.50, 0.51, and 0.54 compositions~\cite{doiron1,sebastian1,audouard1,riggs1,sebastian2} (see Appendix Fig.~\ref{beta1} for an example of the higher $\beta$-frequency observed in current measurements on the $x=$0.54 doping), we focus on the $\alpha$ pocket of carriers that shows the most prominent quantum oscillations in all measured compositions. Figure~\ref{oscillations1} shows examples of quantum oscillations we measure using contactless methods, where changes in the resistivity are reflected as a shift in resonance frequency ($\Delta f$) of an oscillator circuit. Measurements are made down to temperatures of ~1~K using a tunnel-diode oscillator in a slowly swept magnet reaching fields of 55.5~T, and a proximity detector in a two-stage magnet system reaching fields of 85~T (see Appendix). The high magnetic fields used here enable access to the evolution of low energy quasiparticle excitations by the suppression of superconductivity (wherein we refer to the zero resistance state); the crossover field into the high magnetic field resistive state ($H_{\rm r}$) is shown as a function of oxygen composition ($x$) and temperature (Fig. 1 inset). The location of $H_{\rm r}$ is close to the irreversibility field $H_{\rm irr}$ determined by torque measurements in our previous work~\cite{sebastian1}.

\begin{figure}
\centering 
\includegraphics*[width=.46\textwidth]{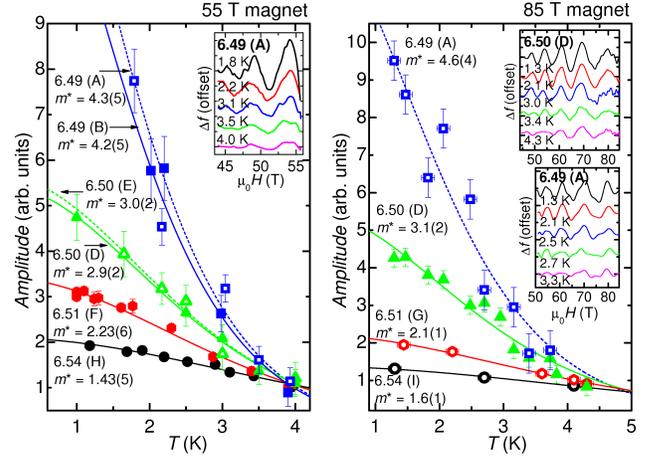}
\caption{Quantum oscillations are measured on nine de-twinned ortho-II ordered YBa$_2$Cu$_3$O$_{6+x}$ single crystals using the contactless conductivity technique in the motor generator-driven 55.5~T magnet and 85~T magnet, with nominal compositions $x=$~0.49 (samples A, B and C), 0.50 (samples D, and E), 0.51 (samples F and G) and 0.54 (samples H, I and J). The figure shows fits of the Lifshitz-Kosevich expression ($a=a_0X/\sinh X$, where $X=2\pi^2m^\ast k_{\rm B}T/\hbar eB$~\cite{shoenberg1}) made to FFT amplitudes of quantum oscillations in seven samples (i.e., A, B, D, E, F, G, H and I) measured in the 55.5~T (a) and 85~T (b) magnet over the field ranges shown in Fig.~\ref{doping1} as a function of temperature ($T$). Amplitudes for each of the dopings are renormalized to coincide at 4~K for comparison purposes. Error bars correspond to the noise floor of the FFT. The inset shows examples of the measured oscillations at several different temperatures for sample 6.49 (A) measured in the 55.5~T magnet, and samples 6.49(A) and 6.50(D) in the 85~T magnet. The same analysis performed over a subset of field ranges is shown in the Appendix.}
\label{mass1}
\end{figure}

\begin{figure}
\centering 
\includegraphics*[width=.4\textwidth]{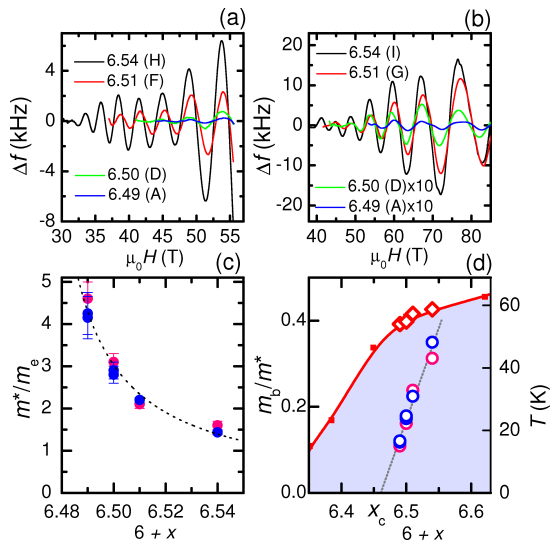}
\caption{Examples of quantum oscillations in each of the samples in Fig.~\ref{oscillations1} at $T~\sim$~1.5 $-$ 1.6~K measured in the 55.5~T (a) and 85~T (b) magnets after background polynomial subtraction. The largest amplitude quantum oscillations of frequency $F_\alpha\sim$~535~T are observed consistently in all compositions. c, Effective mass $m^\ast$  of the quantum oscillations (where $m_{\rm e}$ refers to the free electron mass) extracted from fits shown in Fig.~\ref{mass1} (blue and pink circles referring to the 55.5 and 85~T magnets respectively) plotted as a function of $x$. d, Inverse many-body mass enhancement ($m_{\rm b}/m^\ast$) and a linear fit shown on the left-hand axis as a function of $x$. The superconducting temperature $T_{\rm c}$ as a function of $x$ shown by dots (see Appendix) together with measured $T_{\rm c}$ values for samples A through J (using a SQUID magnetometer) shown by diamonds on the right-hand axis.}
\label{doping1}
\end{figure}

The quasiparticle effective masses ($m^\ast$) are extracted by performing a Lifshitz-Kosevich fit to the temperature dependent amplitude of the observed quantum oscillations~\cite{shoenberg1} (shown in Fig.~\ref{mass1}). Our key experimental finding is that $m^\ast$ exhibits a steep upturn in samples of progressively lower oxygen concentration $x$ (Fig.~\ref{doping1}). The masses are independent within fit uncertainties of sample (different crystals of the same composition), magnet system (sweep rate), magnetic field range, distance from the irreversibility field, and experimental setup~(see Appendix). The decrease in quantum oscillation amplitude with deoxygenation beyond that expected for the increase in mass indicates a Landau level broadening$-$ associated either with increased oxygen disorder, a stronger pairing potential or an increased probability of scattering reflecting the increase in $m^\ast$. In contrast to the striking increase in $m^\ast$, the cross-sectional area of the pocket $A_k=(2\pi e/\hbar)F$ (where $F$ is the observed quantum oscillation frequency in reciprocal magnetic field $1/B\approx 1/\mu_0H$~\cite{shoenberg1}) shows a comparatively weak dependence on $x$~(see Appendix). 

\begin{figure}
\centering 
\includegraphics*[width=.4\textwidth]{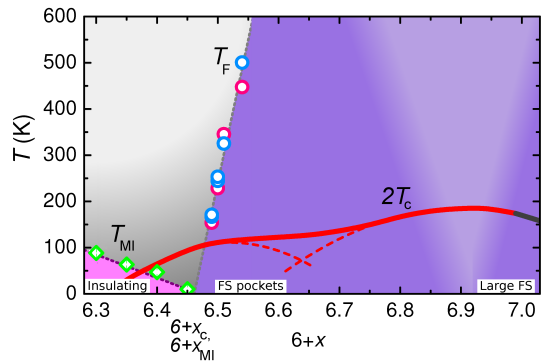}
\caption{Diagram in which the Fermi temperature $T_{\rm F}$ is extracted from the measured values of $m^\ast$ and $F$ (see text), with blue and pink circles referring to data taken in the 55.5~T and 85~T magnets respectively. Green diamonds represent the metal-insulator transition temperature $T_{MI}$ as described in the text (extracted from Refs.~\cite{taillefer1,ando1}). Fits to $T_{\rm F}$ and $T_{\rm MI}$ are represented by dotted lines and a sharp demarcation between different coloured regions. The intercepts $x_{\rm c}$ and $x_{\rm MI}$ refer to the extrapolations of $T_{\rm F}$ and $T_{\rm MI}$ to $T=$~0, indicating the existence of a putative quantum critical point. The solid red line depicts twice the superconducting temperature $T_{\rm c}$ of YBa$_2$Cu$_3$O$_{6+x}$ at zero field (from the Appendix), while the dark grey line represents its extrapolation to notional higher dopings (based on an analogy with Sr$_{2-x}$La$_x$CuO$_4$). The fuzzy conical region centred at optimal doping represents another quantum critical region postulated to occur where the Fermi surface consisting of small pockets transforms to a large Fermi surface (see text). The red dashed curves represent a possible division of the superconducting dome into two intersecting subdomes overlying two distinct critical points: $x_{\rm c}$, and a postulated critical point at optimal doping.}
\label{diagram1}
\end{figure}

A tuning-driven divergence in $m^\ast$ is identified by a collapse in the inverse many-body mass enhancement ($m_{\rm b}/m^\ast$, where $m_{\rm b}$ is the band mass), and hence Fermi temperature ($T_{\rm F}$) to zero at a critical value of the tuning parameter~\cite{coleman1}. Figure 3d shows the ratio $m_{\rm b}/m^\ast$ as a function of $x$ in YBa$_2$Cu$_3$O$_{6+x}$ ($m_{\rm b}\approx$~0.5~$m_{\rm e}$ is estimated from conventional band theory~\cite{sebastian1} and is assumed to remain constant for the incremental changes in $x$ accessed, given the largely unchanged pocket area), and Fig.~\ref{diagram1} shows the inferred Fermi temperature $T_{\rm F}=\hbar eF/m^\ast k_{\rm B}$. A precipitous linear drop in these quantities is seen with reduced oxygen concentration, presaging their vanishing in the vicinity of a critical doping $x_{\rm c}$. Linear interpolation yields $x_{\rm c}\approx$~0.46 as the location of a putative quantum critical point beneath the superconducting dome in YBa$_2$Cu$_3$O$_{6+x}$ (seen from Fig.~\ref{doping1}d).

Critical behaviour tuned by doping rather than by magnetic field is evidenced by the absence of a discernible magnetic field dependence of either $F$ or $m^\ast$ in the range $\mu_0H\sim$26 to 85~T. Thermal conductivity measurements in zero field on YBa$_2$Cu$_3$O$_{6+x}$, also show a notable drop for $x<x_{\rm c}$~\cite{sun1} (see Appendix), and $\mu$sr measurements at zero field reveal an abrupt change in $\mu$sr line shape below $x_c$~\cite{sonier1}. Intriguingly, the critical doping $x_{\rm c}$ is located at the same region of doping where the postulated crossover from metallic to insulating behaviour of the normal carriers~\cite{sun1,sonier1,taillefer1,li1,ando1} onsets, characterized by a low temperature logarithmic divergence in the resistivity~\cite{taillefer1,li1,ando1}. We extract in Figure~\ref{diagram1} the doping-dependence (for $x<x_{\rm c}$) of the metal-insulator transition (or crossover) temperature ($T_{\rm MI}$) from the in-plane resistivity data reported in Refs.~\cite{taillefer1,ando1}. Here $T_{\rm MI}$ denotes the temperature at which the zero (high) field resistivity reaches its lowest value before logarithmically diverging at low temperatures. From Fig.~\ref{diagram1}, we find that with increasing doping, $T_{\rm MI}$ collapses linearly toward $x_{\rm MI}\approx$~0.46, which denotes the zero temperature metal-insulator transition. The coincidence of $x_{\rm MI}$ and $x_{\rm c}$ at $T=$~0 signals an association of the experimentally observed collapse in Fermi temperature and divergent effective mass with a zero (low) temperature continuous metal-insulator transition. The lack of saturation in the logarithmically diverging resistivity for $x\leq x_{\rm c}$~\cite{taillefer1,li1,ando1} indicates that the metal-insulator QCP demarcates a sharp transformation of the entire body of conduction electrons from small Fermi surface pocket to insulating regime in YBa$_2$Cu$_3$O$_{6+x}$.

Although it was previously considered that disorder (e.g. weak localisation) or band depopulation drives insulating behaviour at $x\leq x_{\rm c}$ in underdoped YBa$_2$Cu$_3$O$_{6+x}$~\cite{rullier1}, the steep upturn we observe in $m^\ast$ unaccompanied by a change in $F$ signals that electron correlations are central in driving the development of insulating behaviour for $x<x_{\rm c}$ in underdoped YBa$_2$Cu$_3$O$_{6+x}$~\cite{brinkman1}. While a correlation-driven metal-insulator transition is not entirely surprising given the proximity to the Mott insulating regime dominated by Coulomb repulsion, the continuous nature of the metal-insulator transition (indicated by the collapse of the resistivity upturn temperature at $x_{\rm MI}$) is unexpected~\cite{imada1}. Furthermore, the location of the observed correlation-driven metal insulator transition away from half filling ($x=$~0) in YBa$_2$Cu$_3$O$_{6+x}$ suggests an alternate theoretical scenario (e.g. Refs.~\cite{comanac1,choy1}) to that originally proposed by Brinkman and Rice~\cite{brinkman1}. One possibility is the interplay of additional interactions other than those considered in the Brinkman-Rice picture. Signatures of magnetic order have been reported on both sides of $x_{\rm c}$; a collapse in spin excitation gap has been reported as $x$ is reduced below $x_{\rm c}$~\cite{li1}, while spin density wave ordering has also been suggested to be responsible for Fermi surface reconstruction at $x>x_{\rm c}$ and $\mu_0\rm H \apprge 30$ T~\cite{sebastian3}. A contender for an order parameter that onsets below $x_{\rm c}$ to drive the continuous metal-insulator transition at finite doping is charge order~-~static charge order has been observed to develop below $x_{\rm c}$ by inelastic neutron scattering measurements~\cite{mook1}. A transformation between local and itinerant magnetism near $x_{\rm c}$ may be indicated, the development of local magnetic moments below $x_{\rm c}$ having been reported from $\mu sr$ experiments~\cite{sonier1}. 

While low temperatures are required for our measurement of quantum oscillations$-$ with their observation requiring $k_{\rm B}T$ to fall well within the Landau level spacing ($\hbar\omega_{\rm c}/k_{\rm B}\sim$~80~K at $B=$~85~T and $x=$~0.54)$-$ the Fermi energy scale ($T_{\rm F}$) associated with the observed Fermi surface pockets extends to energies greatly exceeding $T_{\rm c}$ away from the QCP. The rapid collapse of this energy scale at $x_{\rm c}$$-$ located under the local maximum (or plateau) of the superconducting dome$-$ mirrors the behaviour seen in strongly correlated f-electron superconductors, in which case a diverging effective mass has been reported at a QCP~\cite{monthoux1} under the superconducting dome maximum, where f-electrons are removed from participation in the Fermi surface volume~\cite{shishido1}.
Finally, we note that another quantum critical point (or extended region of criticality) is postulated to occur near optimal doping~\cite{broun1,tallon1,chakravarty1,kivelson1,varma1,zaanen1,sachdev1} (i.e. the maximum of the upper superconducting dome in YBa$_2$Cu$_3$O$_{6+x}$) where the small Fermi surface pockets in YBa$_2$Cu$_3$O$_{6+x}$~\cite{doiron1,sebastian1} are expected to evolve into a large Fermi surfaceÑ  recently observed in Tl$_2$Ba$_2$CuO$_{6+\delta}$~\cite{vignolle1}. An intriguing possibility to consider therefore is the existence of two intersecting superconducting domes in high $T_{\rm c}$ cuprates$-$ perhaps similar to the seminal heavy fermion superconductor CeCu$_2$Si$_2$~\cite{gegenwart1}$-$ where each of the superconducting subdomes is centred at a distinct critical Fermi surface instability.

\section{Appendix}
{\it $T_{\rm c}$ and sample compositions}: The $T_{\rm c}$ curves plotted in Figs. 3d and 4 are taken from Ref.~\cite{liang1}. Non oxygen-ordered samples such as those measured in Ref.~\cite{sun1} were previously reported to have a slightly different $T_{\rm c}$ versus $x^\prime$ dependence to those in Ref.~\cite{liang1}, with a putative metal-insulator transition reported to occur at $x_{\rm MI}^\prime~\sim$~0.55. In this work, for accurate comparison $T_{\rm c}$ values are used as a means of renormalizing doping values ($x$) of ortho-II ordered and oxygen disordered samples grown by different methods (following Li {\it et al.} in Ref.~\cite{li1}). Using renormalized dopings to compare Ref.~\cite{sun1}, then $x_{\rm MI}^\prime$ is equivalent to $x_{\rm MI}\sim$~0.47 for the current samples$-$ close to the extrapolated value shown in Fig. 4.
 
\begin{figure}
\centering 
\includegraphics*[width=.36\textwidth]{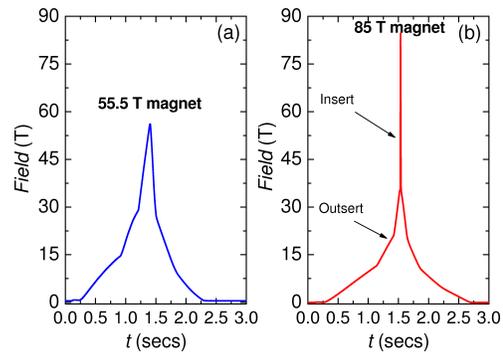}
\caption{Pulse waveform for each of the 55.5 T and 85 T magnets used. The slow sweep rate of the 55.5 T magnet and up to 36 T in the 85T magnet is due to magnetic field generation by a motor generator. The remaining field in the 85T magnet is provided by a capacitor bank.}
\label{magnet1}
\end{figure}

{\it Magnet systems used in the experiments}: Two different magnet systems are used to perform the experiments. Experiments extending to 55.5 T in magnetic field are conducted in a motor-generator driven magnet with a slower sweep rate and longer pulse length (magnetic field versus time profile shown in Fig.~\ref{magnet1}a) than capacitor bank-driven pulsed magnets. Contactless conductivity measurements performed in this magnet use a tunnel diode oscillator circuit with a resonance frequency of $\sim$~46 MHz~\cite{coffey1}. For the experiments conducted in magnetic fields extending to 85 T, an``outsert" magnet powered by the motor-generator is swept slowly to $\sim$~36 T, with the remaining magnetic field provided by a capacitor bank-driven ``insert" magnet (magnetic field versus time profile shown in Fig.~\ref{magnet1}b). For the experiments performed in this magnet, the contactless conductivity measurements use a proximity detector circuit resonating at $\sim$~22 MHz~\cite{altarawneh1}.

A slow ramp rate of the magnetic field is important to reduce the effects of flux dissipation heating. While experiments up to 55.5~T retain a slow ramp rate throughout the pulse, flux dissipation heating in experiments up to 85~T were minimised due to the slow ramp rate up to 37~T in which region the critical current for vortex pinning is expected to be largest. For both these magnet systems, different cryostats, measurement probes and thermometers were used in addition to different contactless conductivity circuits for detection.

{\it Temperature control and measurement}: The increase in $H_{\rm r}$ with decreasing $T$ in the inset to Fig. 1 provides an in-situ secondary confirmation of the sample temperature: the close correspondence between values for $H_{\rm r}$ extracted from static magnetic field measurements and those extracted using motor-generator-controlled magnetic fields demonstrates that the sample is well-coupled to the liquid cryogen, minimising heating due to vortex motion during these measurements.

The effects of flux dissipation heating are minimised by ensuring immersion of the samples in liquid $^4$He throughout. For temperatures (measured using a calibrated thermometer close to the sample) above 2.17~K, the $^4$He is re-pressured by back filling with $^4$He gas after pumping to ensure continued immersion of the sample in the liquid cryogen during the application of the magnetic field. The resistive crossover is seen to be reproduced between rising and falling field. 

\begin{figure}
\centering 
\includegraphics*[width=.4\textwidth]{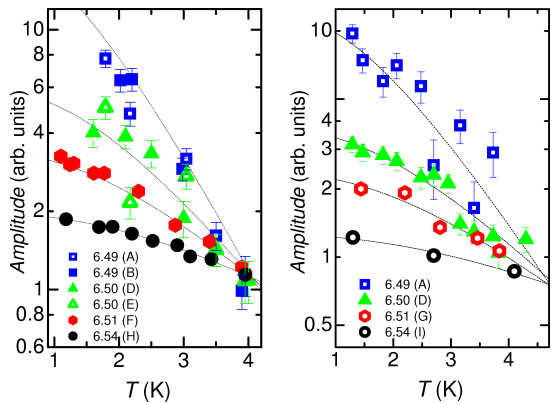}
\caption{FFT amplitudes of quantum oscillations over the limited field range 44 - 55.5 T (measured in the 55.5 T magnet) and the highest field range 60 - 85 T (measured in the 85 T magnet) for all dopings as a function of temperature ($T$). Amplitudes of each of the dopings have been renormalized as in Fig. 2 to coincide at high temperatures.  Error bars correspond to the noise floor of the FFT.}
\label{limitedfft1}
\end{figure}

{\it Extended temperature-dependent amplitudes analysis}: Temperature dependent amplitudes are shown in Fig.~\ref{limitedfft1} over an identical restricted field range 44$-$55.5~T (measured in the 55.5~T magnet) and over the highest field range 60$-$85~T (measured in the 85~T magnet) for all dopings $x$. Similarity with the temperature dependence of amplitude extracted over an extended field range in Fig. 2 indicates that the measured effective masses shown in Figs. 2 and 3, and used to infer the value of $T_{\rm F}$ in Fig. 4 are independent (to within the quoted error bar) of the magnet system, distance from irreversibility field, and magnetic field interval over which it is extracted.

\begin{figure}
\centering 
\includegraphics*[width=.2\textwidth]{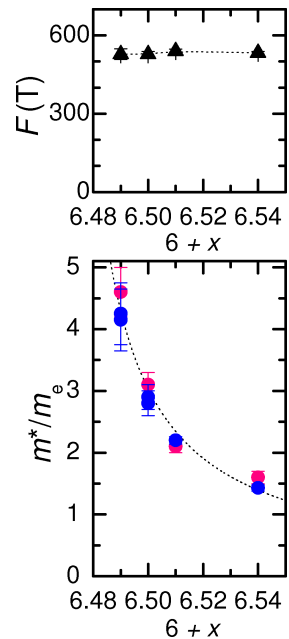}
\caption{(a) Frequency of quantum oscillations corresponding to the $\alpha$ pocket ($F_{\alpha}$) measured as a function of doping in YBa$_2$Cu$_3$O$_{6+x}$. (b) Effective mass   of the quantum oscillations (where $m_e$ refers to the free electron mass) extracted from fits in Fig. 2 (blue and pink circles referring to the 55.5 and 85 T magnets respectively) plotted versus doping.}
\label{limitedmass1}
\end{figure}

{\it Doping-dependent frequency analysis}: Frequencies corresponding to the $\alpha$ pocket were determined by Fourier analysis and by fits of the quantum oscillations in Fig. 2a,b to $A=A_0\cos (2\pi F_\alpha/B+\phi)\exp(-\gamma/B)$ . The frequency can be seen to be largely independent of doping in Fig.~\ref{limitedmass1}a, in contrast to the sharp upturn in effective mass seen in Fig.~\ref{limitedmass1}b.

\begin{figure}
\centering 
\includegraphics*[width=.3\textwidth]{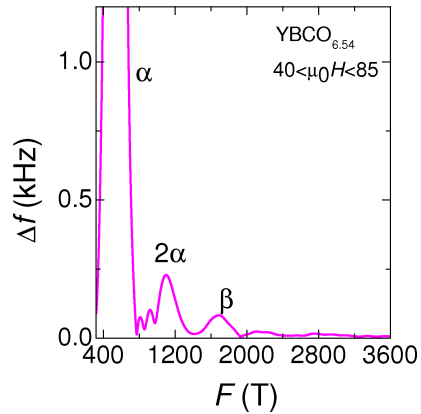}
\caption{FFT of quantum oscillations measured on a single crystal of YBa$_2$Cu$_3$O$_{6.54}$ using the PDO technique in the 85T magnet, revealing the frequency $F_{\beta}~\approx~$1690~$\pm$~20~T, which is 3.14(5) times as large and has an amplitude 0.013 times as small as the most prominent frequency $F_{\alpha}~\approx~$538~$\pm$~5~T.}
\label{beta1}
\end{figure}

{\it The higher $\beta$ quantum oscillation frequency}: The higher $\beta$ frequency has been observed in a subset of measured samples using magnetic torque, contactless conductivity using the Tunnel Diode Oscillator (TDO), contactless conductivity using the Proximity Detector Oscillator (PDO), and specific heat measurements~\cite{sebastian1,riggs1,sebastian2} . An example Fourier transform of the oscillations showing the $\beta$ frequency $F_{\beta}~\sim~1690 \pm 20$~T from sample I of doping $x=$0.54 measured in the 85 T magnet is shown in Fig.~\ref{beta1}.
\\

This work is supported by US Department of Energy, the National Science Foundation and the State of Florida. The authors thank P. B. Littlewood for theoretical input, M. Gordon, A. Paris, D. Rickel, D. Roybal, and C. Swenson for technical assistance, S. Beniwal, G. S. Boebinger, B. Ramshaw, S. C. Riggs, and M. L. Sutherland for discussions.


\newpage

\begin{thebibliography}{99}

\bibitem{millis1} A.~J.~Millis, Phys. Rev. B {\bf 48}, 7183 (1993).

\bibitem{gegenwart1} P.~Gegenwart, Q.~Si, F.~Steglich, Nature Phys. {\bf 4}, 186 (2008).

\bibitem{coleman1} P.~Coleman {\it et al.}, J. Phys.: Cond. Matt. {\bf 13}, R723 (2001).

\bibitem{shishido1} H.~Shishido {\it et al.}, J. Phys. Soc. Japan {\bf 74}, 1103 (2005).

\bibitem{monthoux1} P.~Monthoux, D.~Pines, G.~G.~Lonzarich, Nature {\bf 450}, 1177 (2007).

\bibitem{imada1} M.~Imada, A.~Fujimori, Y.~Tokura, Rev. Mod. Phys. {\bf 70}, 1039 (1998).

\bibitem{sun1} X.~F.~Sun, K.~Segawa, Y.~Ando, Phys. Rev. Lett. {\bf 93}, 107001 (2004). 

\bibitem{sonier1} J.~E.~Sonier {\it et al.}, Phys. Rev. B {\bf 76}, 064522 (2007).

\bibitem{taillefer1} L.~Taillefer 2006 APS March Meeting Abstract: P2.00004.

\bibitem{li1} S.~Li {\it et al.}, Phys. Rev. B {\bf 77}, 014523 (2008).

\bibitem{ando1} Y.~Ando {\it et al.}, Phys. Rev. Lett {\bf 93}, 267001 (2004). 

\bibitem{broun1} D.~Broun, Nature Phys. {\bf 4}, 170 (2008).

\bibitem{tallon1} J.~L.~Tallon, J.~W.~Loram, Physica C {\bf 349}, 53 (2001).

\bibitem{chakravarty1} S.~Chakravarty {\it et al.}, Phys. Rev. B {\bf 63}, 094503 (2001).

\bibitem{kivelson1} S.~A.~Kivelson, E.~Fradkin, V.~J.~Emery, Nature {\bf 393}, 550 (1998).

\bibitem{varma1} C.~M.~Varma, Phys. Rev. Lett. {\bf 83}, 3538 (1999). 

\bibitem{zaanen1} J.~Zaanen {\it et al.}, Nature Phys. {\bf 2}, 138 (2006).

\bibitem{sachdev1} S.~Sachdev, $<$http://arXiv.org/abs/0907.0008v4 (2008)$>$.

\bibitem{rullier1} F.~Rullier-Albenaque, {\it et al.}, Europhys. Lett. {\bf 81}, 37008 (2008).

\bibitem{brinkman1} W.~F.~Brinkman, T.~M.~Rice, Phys. Rev. B {\bf 2}, 4302 (1970).

\bibitem{doiron1} N.~Doiron-Leyraud {\it et al.}, Nature {\bf 447}, 565 (2007).

\bibitem{sebastian1} S.~E.~Sebastian {\it et al.}, Nature {\bf 454}, 200 (2008).

\bibitem{audouard1} A.~Audouard {\it et al.}, $<$http://arXiv.org/abs/0812.0458v1 (2008)$>$.

\bibitem{riggs1} S.~C.~Riggs {\it et al.}, 2009 APS March Meeting Abstract: L33.00004.

\bibitem{sebastian2} S.~E.~Sebastian {\it et al.} (unpublished 2009).

\bibitem{shoenberg1} D.~Shoenberg, Magnetic oscillations in metals (Cambridge University Press, Cambridge 1984).

\bibitem{comanac1} A.~Comanac {\it et al.}, Nature Phys. {\bf 4}, 287 (2008). 

\bibitem{choy1} Y.-O.~Choy, P.~Phillips, Phys. Rev. Lett. {\bf 95}, 196405 (2005).


\bibitem{sebastian3} S.~E.~Sebastian $<$http://arXiv:0907.2958v1 (2009)$>$.

\bibitem{mook1} H.~A.~Mook, P.~Dai, and F.~Dogan, Phys. Rev. Lett. {\bf 88}, 097004-1 (2002).

\bibitem{vignolle1} B.~Vignolle {\it et al.}, Nature {\bf 455}, 952 (2008).


\bibitem{liang1} R. Liang, D. A. Bonn, W. N. Hardy, Phys. Rev. B {\bf 73}, 18505(R) (2006).
\bibitem{coffey1} T. Coffey {\it et al.}, Rev. Sci. Instr. {\bf 71}, 4600 (2000).
\bibitem{altarawneh1} M. M. Altarawneh, C. H. Mielke, J. S. Brooks, Rev. Sci. Instr. {\bf 80}, 4600 (2009).

\end{thebibliography}
\end{document}